\documentclass[11pt,twoside]{article}
\usepackage{./asp2014}

\aspSuppressVolSlug
\resetcounters

\begin{document}

\title{Galactic Archeology -- requirements on survey spectrographs}
\author{Sofia Feltzing,$^1$
\affil{$^1$Lund Observatory, Department of Astronomy and Theoretical
Physics, Lund, Sweden; \email{sofia@astro.lu.se}}}

\paperauthor{Sample~Author1}{sofia@astro.lu.se}{0000-0002-7539-1638}{Lund Observatory}{Department of Astronomy and Theoretical Physics}{Lund}{}{SE-221 00}{Sweden}

\begin{abstract}
Galactic Archeology is about exploring the Milky Way as a galaxy by, mainly, using its (old)
stars as tracers of past events and thus figure out the formation and evolution
of our Galaxy. I will briefly outline some of the 
key scientific aspects of Galactic Archeology and then discuss the associated instrumentations. 
Gaia will forever change the way we approach this
subject. However, Gaia on its own is not enough. Ground-based complementary spectroscopy 
is necessary to obtain full phase-space information and elemental abundances for stars
fainter than the top few percent of the bright part of the Gaia catalogue. 
I will review the requirement on instrumentation for Gaia follow-up that
Galactic Archeology sets. In particular, I will discuss the requirements on radial velocity and elemental
abundance determination, including a brief look at potential pit-falls in the abundance analysis (e.g.,
NLTE, atomic diffusion). This contribution also provides a non-exhaustive comparison of the 
various current and future spectrographs for Galactic Archeology. 
Finally, I will discuss the needs for astrophysical 
calibrations for the surveys and inter-survey calibrations. 
\end{abstract}

\section{Introduction}
\label{sect:intro}

We can approach the problem of formation and evolution of galaxies  in several ways: by observing the properties of galaxies
 back in time, by simulating the formation and evolution of  structure in the universe, or by looking in our own backyard. 
 The first approach has seen impressive progress in the last decades using ever-deeper observations to probe the
  properties of the first galaxies and to charter their evolution till today \citep{2014ARA&A..52..291C}. The second 
  approach has been equally successful. Large simulations have shown that the Cold Dark Matter theory ($\Lambda$CDM) 
  is remarkably able to correctly predict the large-scale structure of the universe as observed today 
  \citep{2005Natur.435..629S,2014MNRAS.444.1518V}. However, it has not (yet) met with equal success on small 
  scales \citep[see, e.g.,][]{2014MNRAS.445..175G}. One reason for this could be that we do not fully understand 
  how baryons influence the processes that lead to actual galaxies \citep{2014MNRAS.444.1518V}. 

Understanding the formation and evolution of galaxies is therefore  foremost about understanding how the baryons are 
distributed. The universe is dominated by dark matter and hence the distribution of baryons is governed by their mutual 
interactions as well as the gravity from the dominant dark matter. Baryons largely reside in stellar disks in galaxies, 
understanding these disks then becomes a key to solving the question of galaxy formation and evolution across the 
ages of the universe.  

The Milky Way is one of billions of galaxies. As a galaxy it is not very remarkable.
In fact its a rather typical galaxy and
 it is the one galaxy we can study in exquisite detail. Therefore, it provides a fundamental test-bed for our theories of galaxy 
formation and evolution \citep{2002ARA&A..40..487F,2013A&ARv..21...61R}. Indeed, it is on these the smallest scales where the firmest 
tests of $\Lambda$CDM can be done \citep[e.g.,][]{2014MNRAS.442.3745M}. 

The data from the astrometric ESA satellite Gaia is set to change the investigation of the Milky Way 
as a galaxy forever.  Gaia is measuring distances and on-sky motions for a billion
stars in the Milky Way and has been 
collecting scientific data since  July 2014. A first data release is planned for 2016, with an expanded release in mid-2017 
\footnote{The data release scenario for Gaia is available at \url{http://www.cosmos.esa.int/web/gaia/release}.}. 

\begin{figure}[t]
\begin{center}
\includegraphics[width=11cm]{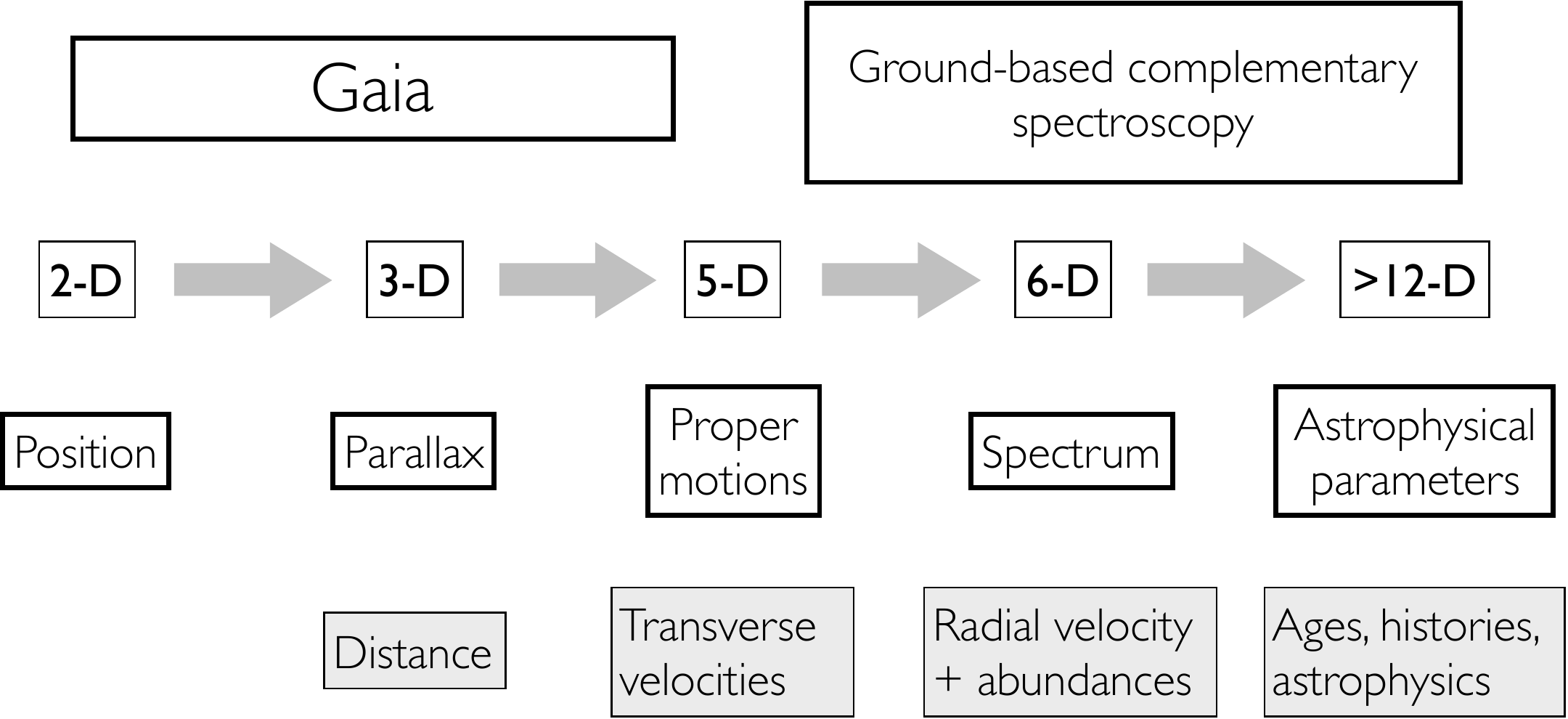}
\end{center}
\vspace{-0.6cm}
\caption{Illustration of how the scientific information is built up as more data is added. The 
 boxes in the third row shows which parameters are possible to determine as 
the dimensionality of the data increases. The fourth row (shaded boxes) show the information
available. Adapted from \citet{2012Msngr.147...25G}.}
\label{fig:gaia12D}
\end{figure}

To make a major breakthrough in understanding the evolution of the Milky Way and its components, detailed information 
beyond positions and velocities (i.e. Gaia) is essential. Such data is provided by a new generation of spectroscopic surveys designed to
obtain high quality spectra for millions of stars: APOGEE \citep{2014ApJS..211...17A}, GALAH \citep{2015MNRAS.449.2604D}, 4MOST \citep{2014SPIE.9147E..0MD}, and WEAVE 
\citep{2010SPIE.7735E..7GB}. Additional spectra of lower spectral resolution will be obtained with instruments
such as PFS \citep{2014PASJ...66R...1T} and DESI \citep{2015AAS...22541304P}. All these spectra will give full 3D motions, elemental abundances and stellar parameters, such that 
ages can be determined for turn-off stars. Asteroseismic data offer the possibility to obtain ages for red giant stars \citep{2013ARA&A..51..353C}. Figure\,\ref{fig:gaia12D} shows how the information about the Milky Way stellar populations is build up as information is added. 

\section{Requirements on the design of a multi-object spectrograph from Galactic Archeology}
\label{sect:req}

When considering the design of a new multi-object spectrograph for Galactic Archeology a number of 
things are very desirable to have. The stellar spectroscopist interested in elemental abundances
in stars and in the stars themselves will be wanting high resolution, large wavelength coverage,
and good sampling of the spectra while those working on the kinematics and dynamics of the 
Milky Way would prefer as many stars as possible with velocity and some abundance information. 
These partly contradictory requirements should then be possible to fit into a design that meets the 
throughput criteria.

{\emph {What do I want to measure and to what precision?}} This can seem as a rather obvious
question, but it is one of the very first questions that must be answered in order to set the requirements
for the multi-object spectrograph. For Galactic Archeology, two measurements are of 
immediate interest: 1) radial velocities; 2) elemental abundances to some precision (and maybe accuracy).

Let's first look at what requirements we may have on the measurement on radial velocities. Most large spectroscopic
surveys are today, as discussed in Sect.\,\ref{sect:intro}, driven to be complementary to Gaia \citep{2008ewg4.rept.....T}. Gaia will deliver parallaxes and proper motions down to $V \sim 20$. The spectra 
taken on board Gaia will deliver radial velocities down to $V \sim 15.2$ for a G2 stars. At that magnitude 
the errors in radial velocities are already large (around 15\,km\,s$^{-1}$)\footnote{As described on the web-page for Gaia science performance. These numbers were read of 
that page on 6 May 2015.  \url{http://www.cosmos.esa.int/web/gaia/science-performance}.}. 
 Thus, there is a direct need to complement Gaia with radial velocities obtained from the ground to get 
 the full 3D motion for a large fraction of the stars down to $V\sim 20$. Such
 radial velocities are readily obtainable and errors should be around $2-3$\,km\,s$^{-1}$ to match the errors
 in Gaia's proper motions all the way down to $V \sim 20$.

\articlefigure[width=10cm]{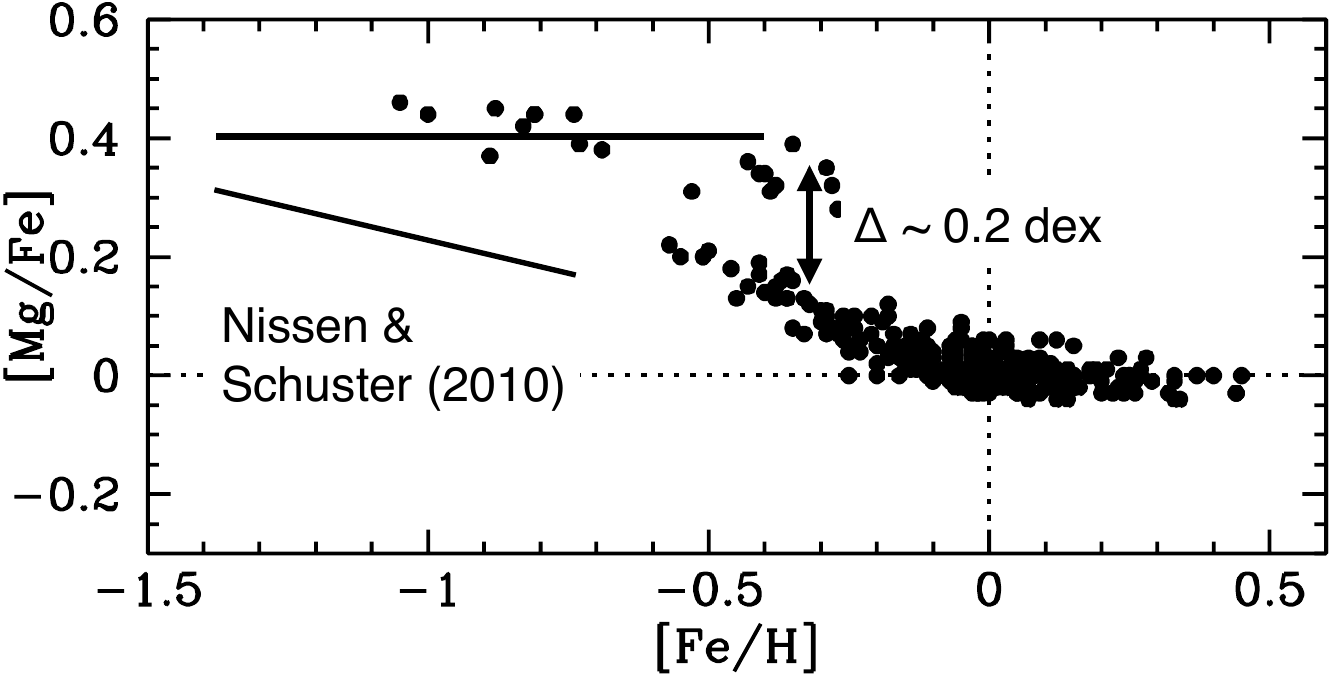}{fig:lines}{[Mg/Fe] as a function of [Fe/H] for stars within 25\,pc of the sun analysed by Klaus Fuhrmann \citep[presented in a series of papers, e.g.,][and here represented by a collection of his  data]{2011MNRAS.414.2893F}. The gap between the upper and the lower trends is about 0.2\,dex. The two straight lines (one horizontal and one slanted) 
delineates the data from \citet{2010A&A...511L..10N}. They study stars in the solar neighbourhood with typical halo kinematics and find 
that some follow the "classical" $\alpha$-enhanced high trend, whilst about half of their stars follow a decreasing 
trend. \emph{Note}: This is not an extension of the thin disk. The stars have clear halo kinematics. }

Secondly, let's  look at what requirement we might want to set from the point of view of elemental abundances. 
Our approach here is to look at well-known data from the solar 
neighbourhood and assume that we \emph{at least} wish to be able to obtain data that can distinguish the
features seen in the solar neighbourhood.  The assumption being that it is rather likely that similarly sized
features will be present also in other parts of the Galaxy. A pertinent example is given by the various data by Klaus Fuhrmann 
\citep[presented in a series of papers, e.g.,][Fig.\,\ref{fig:lines}]{2011MNRAS.414.2893F}.
As we can see, the difference between his two trends is about 0.2\,dex. Another example in the solar neighbourhood is 
provided by \citet{2010A&A...511L..10N}.  They study stars in the solar neighbourhood with typical halo kinematics and find 
that some follow the "classical" $\alpha$-enhanced high trend, whilst about half of their stars follow a decreasing 
trend (Fig.\,\ref{fig:lines}). Again, the typical size of the separation is about 0.2\,dex. These are a very secure result and by now reproduced
in other studies \citep[e.g.,][]{2014A&A...562A..71B}.

Thus, a brief tour of the solar neighbourhood shows us that we need to be able to resolve differences in elemental
abundances of about 0.2\,dex in a statistically significant manner. This can be achieved by either observing a large
number of stars with errors not much smaller than the separation, or by observing a smaller number of stars
at much higher precision. This is discussed in some detail in \citet{2013A&A...553A..94L}.

\subsection{Resolving power and signal-to-noise ratio}

Radial velocities in a big spectroscopic survey are best measured from lines that remain strong and well defined in a large 
variety of stellar types.  Such lines include the three Ca\,{\sc ii} triplet lines in the NIR (at 849.8, 854.2, and 866.2\,nm) and the three so called 
Mgb lines (at 516.7, 517.2, and 518.3\,nm). 
Radial velocities are readily measured from these strong lines \citep[e.g., in RAVE,][]{2013AJ....146..134K}. An increased resolving power  improves the precision in the measured radial velocities. In general $\sigma_{\rm RV} \propto R^{-3/2}$ (technical note by M. Irwin for WFMOS and P. Bonifacio, priv. comm.), this means that the precision in radial velocities for a resolving power of 7\,500 is a factor 0.76 smaller than for a resolving power of 5\,000 (keeping the SNR {\AA}$^{-1}$ and wavelength coverage fixed). Such resolving power is
also relatively well matched to what is needed for extra-galactic science \citep{2014PASJ...66R...1T}. Instruments like DESI and PSF are
mainly designed for the extra-galactic science and have thus lower resolution, whilst, e.g., the deign of the low-resolution
spectrographs in 4MOST is primarily driven by the need to complement Gaia and hence has a goal of 7\,500.

Good elemental abundances with high precision are typically recovered from spectra of high to very high resolving power (say 
40 000 to 100 000) accompanied with high signal-to-noise ratios, typically $>250$ per reduced pixel, and in some cases
significantly higher \citep[see, e.g.,][for two examples]{2014ApJ...791...14M,2014A&A...562A..71B}. Such high resolving power would come at the expense of very short wavelength coverage
of the spectrograph. Most current and future multi-object spectrographs have therefore made a compromise and settled for a lower resolving power but with a larger wavelength coverage \citep[e.g., WEAVE, HERMES/GALAH;][respectively]{2010SPIE.7735E..7GB,2015MNRAS.449.2604D}. That spectra with an $R\sim 20\,000$ are capable of delivering good elemental abundances has been shown 
by the Gaia-ESO Survey analysis of the FLAMES/GIRAFFE spectra and early results from the GALAH survey.

Signal-to-noise and resolving power can  be traded off against each other. If we consider a single line for which we wish to
derive an elemental abundance from it can be shown that the relative error in the retrieved equivalent width, $W_{\lambda}$,
is $\propto R^{-1/2}$ \citep{1992ESOC...40...17G}. This error is directly proportional to the error in the elemental abundance derived
for this line. This means, e.g., that by going between $R\sim 16\,000$ and $\sim 23\,000$
the error in the retrieved $W_{\lambda}$ (and hence in the error for the derived elemental abundance) 
decreases by a factor 0.834, everything else being equal. Thus, a lower resolving power
can (to some extent) be compensated by a longer integration time resulting in a higher signal-to-noise ratio.

\subsection{Wavelength coverage}

However, how to reach a suitable compromise on wavelength coverage and resolving power? This is not a very easy 
question and here we will only be able to outline some aspects of how to proceed from a science requirement to
a requirement on wavelength coverage (and indirectly resolving power) for a spectrograph. 

Figure\,\ref{fig:lambda_sum} and Table\,\ref{tab:spect} show compilations of the wavelength coverage for a number
of current and planned survey facilities operating in the optical. In addition two spectrographs, APOGEE and MOONS, 
have opted for NIR
wavelength coverage in order to be able to study stars in the Galactic plane where interstellar extinction is heavy
\citep{2014ApJS..211...17A,2014SPIE.9147E..0NC}. We do not discuss these two instruments any further in
this section.

\begin{table}[!h]
\addtocounter{footnote}{-1}
\caption{Summary of wavelength coverage and resolving power ($R$) for spectrographs
with low resolving power, including the low resolution (LR) spectrographs for WEAVE 
and 4MOST. Please note that some of these values are subject to change as the 
final design of the spectrographs is developed (DESI, WEAVE, and 4MOST). WEAVE
and 4MOST both have the goal of 7\,500 in resolving power.}
\smallskip
\begin{center}
{\small
\begin{tabular}{lllrll}  
\tableline
\noalign{\smallskip}
Spectrograph & $\lambda$-range & $R$ & \# fibres & Reference/Website\\
& [nm] & & in LR  \\
\noalign{\smallskip}
\tableline
\noalign{\smallskip}
SEGUE & 380 -- 920 & 1\,800 & 640& \citet{2000AJ....120.1579Y} \\
RAVE & 841 -- 879.5 & 7\,500 & 150 & \url{https://www.rave-survey.org/}\\
LAMOST & 370 --  900 & 1\,800 & 4\,000 & \url{http://www.lamost.org}\\
PFS & 380 -- 1260 & & 2\,400 & \citet{2014PASJ...66R...1T}\\
       & 380 -- 670 & 1\,900 \\
       & 650 -- 1000 & 2\,400 \\
       & 970 -- 1260 & 3\,500\\
DESI & 360 -- 980 & 4\,000\protect\footnotemark & 5\,000 & \url{http://desi.lbl.gov} \\
WEAVE LR &  366  -- 959& 5\,000 & 1\,000 & \citet{2014SPIE.9147E..0LD}\\
4MOST LR & 390 -- 930 & 5\,000 & 1\,600 & \citet{2014SPIE.9147E..0MD}\\
\noalign{\smallskip}
\tableline
\end{tabular}
}
\end{center}
\label{tab:spect}
\end{table}
\footnotetext{$R$ as measured in the infra-red.}

For the low resolution spectrographs the wavelength coverage is rather similar
for all the instruments. 
For Galactic Archeology the blue end is of a 
specific interest as very metal-poor stars have most
of their spectral features below about 460\,nm  \citep[see, e.g.,][for a definition of very metal-poor]{2013pss5.book...55F}. The region can be used to find and/or characterise the overall properties
of the very metal-poor stars. In addition, some interesting elemental abundances, e.g., Sr can be derived 
\citep[see, e.g.,][]{2007ApJ...667.1185L}. Hence, even a spectrograph with relatively low resolving power, 
e.g., SEGUE and DESI, can be 
a highly useful tool in Galactic Archeology. 
To obtain radial velocities the Mgb triplet lines and the Ca\,{\sc ii} NIR triplet lines are very useful as they
are present in essentially all late type stars and across all metallicities. Hence, from the point of view of Galactic 
Archeology a first requirement on the wavelength coverage of a low-resolution spectrograph must include these 
lines. This is not a very strong requirement. For the characterisation of metal-poor stars inclusion of the region
below 400\,nm with the Ca\,H{\sc \&}K lines is necessary.

For the high-resolution spectrographs (typically, $R\sim 20\,000$ for these instruments) the wavelength coverage
varies. Sometimes considerably. The now more than ten year old FLAMES/GIRAFFE instrument has full coverage 
in the high-resolution mode of the optical spectral range, but cut up into 22 set-ups, each with an effective 
wavelength range of $\sim \lambda$/22 \citep[][and Fig.\,\ref{fig:lambda_sum}]{2002Msngr.110....1P}. For the Gaia-ESO Survey two regions were chosen for
the Milky Way science \citep[one region centred at 548.8\,nm and one centred on the Ca\,{\sc ii} triplet lines in the NIR,][]{2012Msngr.147...25G}. 
The Ca\,{\sc ii} triplet lines in the NIR was specifically chosen to enable determination of surface gravity. As the total wavelength coverage
is limited it is important to ensure that  gravity and temperature sensitive spectral features are available
in the wavelength region. 

\articlefigure[width=6.5cm,angle=-90]{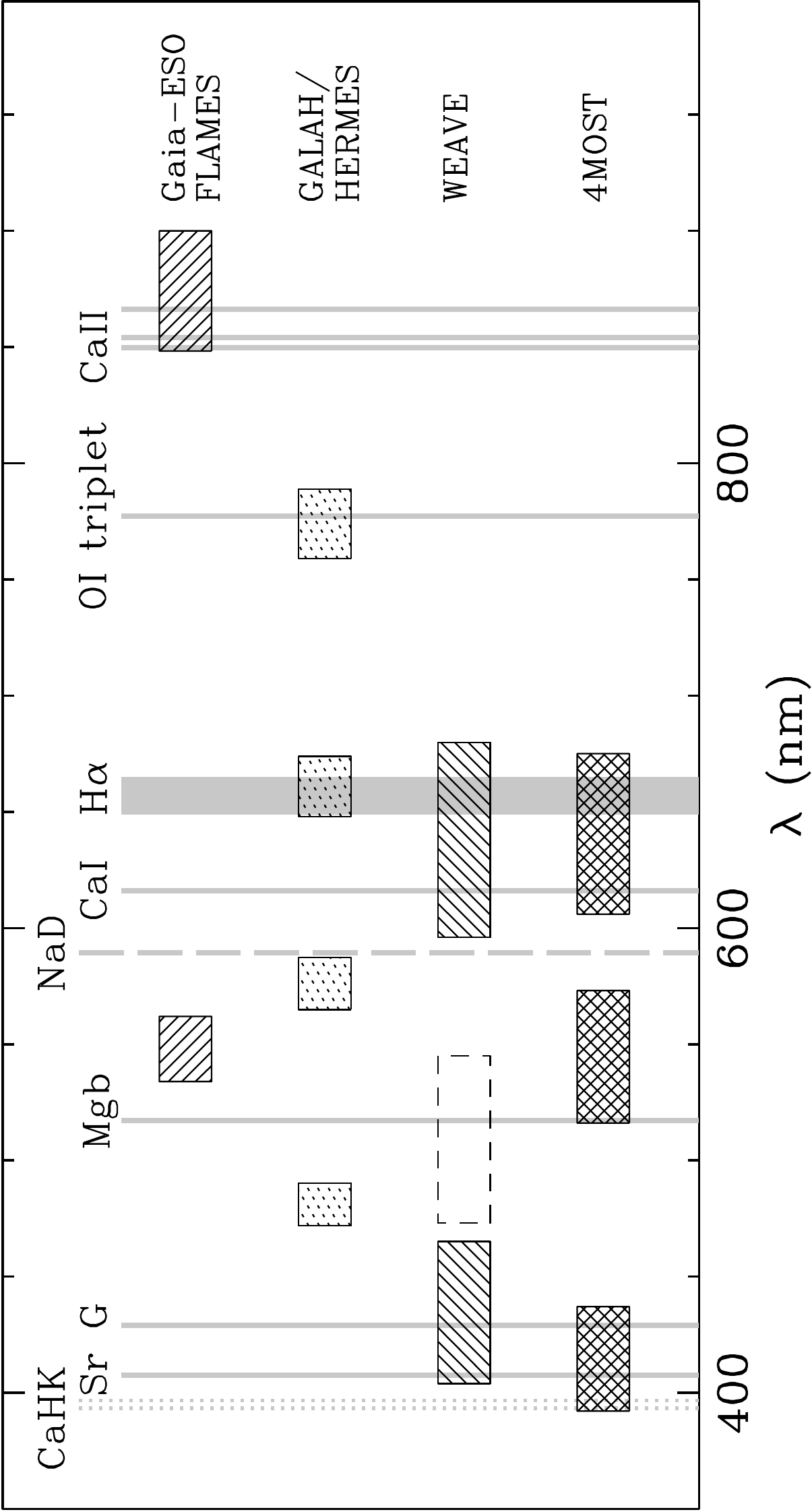}{fig:lambda_sum}{Wavelength coverage in the 
optical for a number of current and future survey instruments  with high resolving power that can be used for Galactic 
Archeology. The compilation is not meant to be exhaustive but illustrates the type of wavelength coverage available
 to do our science. For WEAVE we also include the  third band that is possible to use instead of the bluest band. 
 This has mainly been designed for extra-galactic studies. A number of interesting spectral features are indicated 
 at the top of the plot, and light grey lines show their position. {\tt G} stands for the G-band. {\tt Mgb}   for 
 the Mg triplet lines, {\tt CaI} for the gravity sensitive line at 616.2\,nm, and {\tt CaII} for the Ca\,{\sc ii} triplet lines 
 in the NIR. The positions of the Ca\,H{\sc \&}K lines are marked by dotted lines
 and the position of the interstellar NaD lines are shown as a dashed line. The width of the line indicating
 the position of the H\,${\alpha}$ line shows the region that needs to be included in order to sample the whole
 wings, including taking a radial velocity of up to 400 km\,s$^{-1}$ of the star into account.}

HERMES, which is used for GALAH,  WEAVE and 4MOST   have made rather different choices of wavelength 
coverage (compare Fig.\,\ref{fig:lambda_sum}). The total wavelength coverage of HERMES is about 1000\,nm  
divided into four roughly equal wavelength ranges, covering lines useful for stellar parameters (e.g., H$_{\alpha}$) 
as well as a multitude of elemental abundances \citep{2015MNRAS.449.2604D}. HERMES is set apart from 
the other upcoming instrument designs by the inclusion of the oxygen triplet lines around 777\,nm. This is a 
unique feature, which will allow the GALAH team to derive oxygen abundances for their whole sample. The 
oxygen triplet is plagued by analysis difficulties but most of these have recently been overcome and hence 
the inclusion of this region, which holds little other information, is highly valuable to Galactic Archeology.

The WEAVE and 4MOST designs have partly been driven by the wish to include studies of very metal-poor 
stars  (i.e., stars with [Fe/H] $<-1$). Such stars have essentially all spectral features that are possible to 
analyse in the blue, below around 450\,nm  \citep[see, e.g.,][for an  example of a complete analysis of a
 very metal-poor stars]{2003ApJ...591..936S}. Of particular interest are the CH-band at [429:432]\,nm, 
 the Ca\,H{\sc \&}K lines, Sr\,{\sc ii}, Ba\,{\sc ii}  \citep[][and Hansen et al. submitted]{2013AN....334..197C}.   
 The final designs have to take into account finite CCD sizes and constraints from the optics. 4MOST have 
 opted for a bluer cut-off end including the Y line at 395.0\,nm and the Ca\,H{\sc \&}K lines at 396.8 and 
 393.3\,nm,  whilst WEAVE has opted for a wider coverage in the blue channel reaching further  in to the 
 red, thus making sure several interesting species, such as Ba\,{\sc ii} at 455.4\,nm, are fully included.  
 Both instruments include the G-band and the bluest Sr\,{\sc ii} (compare Fig.\,\ref{fig:lambda_sum}). 
  It is a difficult task to select this wavelength region as almost every line represents a unique opportunity. 
  Thus, there is hardly any perfect tradeoff.  A paper discussing
 the trade-offs for the blue channel is submitted (Hansen et al.).

The reddest channel in WEAVE and 4MOST are quite similar, with WEAVE being the wider one. Both include
the H$_{\alpha}$ line (sensitive to effective temperature) and the Ca\,{\sc i} line at 616.2\,nm
 \citep[sensitive to surface gravity,][]{1988A&A...190..148E}. In addition, 4MOST has a green channel that is 
specifically designed to ensure the inclusion of a) additional features sensitive to surface gravity (Mgb lines),
b) that the number of Fe\,{\sc ii} lines suitable for analysis in disk turn-off dwarf and red clump stars
 included in the green and red channel is at least ten to get a separate test on surface gravity, and c) similarly,
  that there are at least 10 Fe\,{\sc i} lines with high and low excitation energies, respectively, 
 included in the combined green and red channels to give an independent measure of effective temperature.
 The full discussion of possible choices of wavelength coverage for 4MOST and how this selection process
 can be quantified leading to an optimised wavelength range 
 will be presented in a forthcoming paper by Ruchti et al. (in prep.).

\section{Analysis of stellar spectra in large surveys -- promises and pitfalls}
\label{sect:analysis}

The possibility to derive a good elemental abundance from a stellar spectrum is hampered by numerous problems; including
departures from Local Thermal Equilibrium (LTE), lack of atomic data, an inability to account for the full 3D structure of the stellar atmosphere. These issues are the subject of many studies. In this brief review we give the reader a few pointers
 to selected studies. A series of recent papers explore the issues of NLTE in 
the determination of iron abundance, stellar parameters, and ages \citep[][]{2012MNRAS.427...27B,2012MNRAS.427...50L,2013MNRAS.429..126R,2013MNRAS.429.3645S}. Differences in abundances
derived in LTE compared to so called averaged 3D plus NLTE models show differences of up to 0.4\,dex
for warmer stars as we move from sub-solar metallicities down to --3\,dex \citep{2013MNRAS.429..126R}. For two recent reviews of NLTE
calculations for other elements see \citet{2014arXiv1403.3088B} and \citet{2014IAUS..298..355M}.  
\citet{2013A&A...557A..26M}  provides details on the
new grids of 3D model atmospheres that are being developed and are now also being adapted in detailed elemental
abundance work. An early application is discussed in \citet{2007A&A...469..687C}. 
Application to large number of stars is becoming feasible, e.g., averaged versions are now included in some abundance pipelines
used in on-going surveys.

Diffusion in the stellar atmosphere is still little explored, but has been shown to have a clear effect on stars in
globular clusters at varying metallicities, such that stars at the turn-off and on the sub-giant branch are depleted
in iron and in titanium \citep[][and references therein]{2013AN....334..114M}. 
Empirically it has been shown that the effects can be as large as 0.2\,dex 
\citep{2007ApJ...671..402K,2013A&A...555A..31G,2014A&A...562A.102O}. However, given that the effects
of diffusion as observed in these clusters, is roughly the same for all elements, a differential approach where
abundance ratios are compared reduced the "error" between turn-off and red giant branch stars to a much
lower level, of the order 0.05\,dex or less.

Such large uncertainties resulting from a lack of knowledge of the physics  might feel discouraging in 
the context of Galactic Archeology, where, as discussed in Sect.\,\ref{sect:req}, we aim at a precision in relative
abundance of less than 0.05 -- 0.1\,dex (depending on the science question). However, given the 
significant progress in recent years and the on-going efforts in understanding the physics inside stars 
and stellar atmospheres the future is bright with real possibilities to overcome the difficulties in
achieving a homogenous analysis that is able to combine data from stars of very different evolutionary
stages and metal-content into a combine picture of what the Milky Way looks like. 

Just as with diffusion, differential studies remains an important way to mitigate some of the errors
introduced by our limited knowledge of the physics involved. By studying only stars with very similar stellar 
parameters it is possible to, ot first order, mitigate issues such as NLTE and lack of knowledge of the structure of
the stellar atmosphere. Examples where this approach has enabled the discovery of interesting 
abundance trends and differences include \citet{2010A&A...511L..10N} and  \citet{2015arXiv150407598N}.

\section{Calibrating Galactic surveys}

Observations of science calibrations have two main purposes for  Galactic Archeology:

\smallskip
\noindent
{\bf Astrophysical calibrators} -- which provide a means to calibrate the spectral analysis in terms of 
stellar parameters, elemental abundances and, potentially, age.

\smallskip
\noindent
{\bf Inter-survey calibrators} -- which provide a means to put the survey onto the same abundance scale
as other Galactic spectroscopic surveys.

\smallskip

In addition we should add {\bf Basic calibrators}, i.e., observations that allow us to get rid of sky lines, telluric absorption
lines and to obtain a stable wavelength solution enabling good radial velocity measurements. 
Basic calibrators are typically dealt with in any observing strategy, regardless of the science goal. Hence,
there are by now well-proven methods to provide a stable velocity solution for the spectra. Correction for
sky and telluric lines is important, even critical, for some of the Galactic Archeology goals (e.g., when an
important line falls in the region effected). Inclusion of a few fast rotators or white dwarfs in each field is 
normally deemed an efficient way to deal with the tellurics. Alternatively, models of the Earth's atmosphere
can be used to correct for this \citep[one recent example and application can be found in][]{2015A&A...576A..78K}. To correct for sky lines, numerous sky fibres are normally 
included in each observation.

\subsection{Astrophysical calibrators}

More interesting, from the point of the analysis of the stars, is the calibrators that allows us to 
ensure that our analysis (from the raw spectra to the final elemental abundances) does a good
job. These are the astrophysical calibrators. Several groups of stars could be used for this. Each
with its own pros and cons.

\vspace{-0.15cm}
\begin{enumerate}

\item Stars with very well-defined properties, such as the Gaia-benchmark stars.
\vspace{-0.15cm}
\item Fields of stars with very high-quality data available for a large number of stars, such as the Kepler field or certain open clusters.
\vspace{-0.15cm}
\item Fields of stars with some useful data available, this includes most globular and open clusters but also fields observed by other surveys.

\end{enumerate}

The Gaia benchmark stars are stars with fundamental parameters, such as effective temperature, derived
from basic observations such as the radius of the star\footnote{
For this set of  stars a library of high-resolution, high signal-to-noise spectra have been prepared \citep{2014A&A...566A..98B}.
These spectra are publicly available and can be  used inside any survey for, e.g., for testing purposes. 
They can be downloaded from \url{http://www.blancocuaresma.com/s/benchmarkstars/}.}.
The list was originally put together for the purpose of 
providing fundamental calibrators for the Gaia astrometric mission, they have also been extensively used in current 
spectroscopic surveys, e.g., in the Gaia-ESO Survey \citep{2014A&A...570A.122S}. A paper listing their 
metallicities have been published in  \citet{2014A&A...564A.133J} and a list of the stellar parameters can be 
found in \citet{2014ASInC..11..159J}.

Several  efforts  to obtain further fundamental measurements of stars are on-going. There are plans to 
further capitalise on such efforts and establish additional fundamental calibrating stars across the 
HR-diagram. Recent results include \citet{2015A&A...575A..26C}.

Another approach might be to combine asteroseismology and various photometric systems \citep{2015ASSP...39...61C}.
Even if stellar parameters derived through such means might not be fully as fundamental as those from direct 
measurements this might still be a highly desirable route to explore in order to enlarge the set of benchmark stars
to ensure a complete coverage of the HR-diagram at a wide range of metallicities.

The stars with well-defined properties, the benchmark stars, are fundamental for providing a sound basis 
for the analysis pipelines. Open and globular clusters serve a similar purpose as they provide stars of the 
same age and well-defined abundances along the evolutionary sequence, providing a unique opportunity 
to check the self-consistency of the pipeline that provides the stellar parameters (but see Sect.\,\ref{sect:analysis}). 

\subsection{Inter-survey calibrators}

Finally, we need observations that makes sure that we can combine the different Galactic surveys. This can be achieved 
with the help of the astrophysical calibrators, if they are numerous enough, or alternatively specific observations could be devoted to this.
The key aspect is, of course, that the stars used for this task must have been observed by each of the surveys that should
be combined. As we wish to ultimately combine all surveys, data that will be most useful as inter-survey calibrators are 
fields of stars along the equator, i.e. fields that can be easily observed with surveys covering either the Northern hemisphere or
the Southern hemisphere of the sky.

Typically, we will think of open and globular clusters for such purposes. Clusters are excellent for testing and validating the 
analysis tools as they include stars at many evolutionary stages and all stars are at the same distance. However, globular 
clusters are not well distributed on the sky. Most are observable from the South only as they cluster around the Galactic Centre. 
Open clusters are spread all over the sky, but with a strong concentration around the Galactic plane. Observing directly in the plane
is difficult for the surveys operating in the optical
 thanks to the high extinction. Hence, it is likely best to select open clusters slightly off the plane. In addition, there are
only a relatively small range of the Galactic plane that is observable from both the Northern and the Southern hemisphere.

On-going and completed surveys have already observed a significant number of globular and open clusters. Future surveys need to consider how to best relate to that data as well as enlarging the possible inter-surveys calibrating fields.
One potential would also be to include regular science fields from other surveys and use these as inter-survey calibrators. 
This might seem a little sub-optimal, but if the main aim is to ensure that a certain type of star looks the same in each of 
the surveys (and not thinking about precis or accurate stellar parameters) this might be a fruitful way forward. The fields observed
in GALAH should be possible to adapt to the deeper surveys, such as 4MOST and WEAVE. To go from the deeper to the shallower
surveys might, on the other hand, prove too time-consuming to be realistic.

\subsection{Asteroseismic fields}

Several Milky Way fields now have asteroseismic observations. These fields have been observed
by CoRoT and Kepler \citep{2009A&A...506..411A,2010Sci...327..977B}, and new fields will be observed in the Kepler II campaigns\footnote{Further links to the campaigns can be found at \url{http://keplerscience.arc.nasa.gov/K2/}.}. In the future PLATO will provide several fields \citep{2014ExA....38..249R}. 

The interesting aspect of the seismic data is that the surface gravity of the star and its age can be determined
with very high precision \citep{2013ARA&A..51..353C}. This is highly valuable information for many purposes and a very exciting prospect
for astrophysical calibrations of Galactic Archeology. Indeed, as these fields have such rich seismic data sets they will be, 
and already are, natural targets for spectroscopic 
surveys. APOGEE is observing the Kepler field and will also look at the Kepler\,II fields, so is GALAH. The 
Gaia-ESO Survey has observed both the inner as well as the outer CoRoT field. Hence, these fields provide a 
very good opportunity for cross-calibrating the spectroscopic surveys as long as the same stars have been observed.

Targets in the Kepler\,II fields are relatively bright and hence accessible to all surveys.
 Prime Kepler\,II targets have $V <14$ (Victor Silva Aguirre, {\it priv.com.}). 

The asteroseismic fields are large and several contains one or more open clusters as they are interesting targets
for the seismic analysis. The fields are typically larger than the FoV of WEAVE or 4MOST. To obtain good cross-calibrations
there is a need to start inter-survey discussions now to ensure that the same actual stars are observed.

\section{Summary and what's next?}

To conclude, instrumentation for Galactic Archeology in the form of massively multiplex spectrographs 
is now being realised on a large scale with massive surveys of both the Southern and the Northern 
hemisphere. These surveys will provide excellent complimentary data to the parallaxes and proper 
motions from Gaia, enabling full phase-space information as well as further information on stellar
parameters and elemental abundances. Ages are crucial for disentangling the formation history of
the Milky Way. For turn-off stars ages can be determined using stellar isochrones, whilst asteroseismology
offers the promise of ages for red giants stars. 

However, the effort to carry out large surveys should be matched by efforts to improve our analysis techniques
and our understanding the physics inside stars and their atmospheres. We also need to spend m
re effort
in establishing benchmark type stars across the whole HR-diagram. Such stars are needed for several
purposes and can be used in analysis methods such as the Cannon \citep{2015arXiv150107604N}.

Going beyond the currently planned survey instruments it is interesting to ask the question: What's next?
RAVE and SEGUE provided relatively large number of targets at low or modest resolution. The next step
has been to increase the resolving power in the instruments, e.g., WEAVE, HERMES, 4MOST, as well
as increasing the number of fibres, e.g., 4MOST and DESI. What is the next desirable step? Is it to provide
even more fibres over an even bigger field? Is it to provide fewer fibres but with increased resolving power? 
Or, would a single slit instrument on a large, $8-10$\,m, telescope be a better way to provide the next
steps in Galactic Archeology?

Many of the current surveys are or will provide great catalogues that need to be followed up at higher 
resolution and/or higher signa-to-noise ratios. On the other hand, Gaia will provide additional, still unknown
targets. Those we can easily think of are stars that appear to cluster in phase-space, e.g., moving groups or 
dispersed stellar clusters. What is the best way to follow-up such targets? These stars are likely not clustered
on the sky. Hence, there appear to be a case for a single slit high-resolution, high signal-to-noise ratio 
spectrograph for survey and Gaia follow-up.


\acknowledgements The author acknowledges  conversations with Mike Irwin, Luca Sbordone, Greg Ruchti,
  Karin Lind, Elisabetta Caffau and Camilla Juul Hansen for instructive discussions on science requirements
  for spectrographs. For prospects of cross-calibrating all the major spectroscopic
surveys the author wish to thank Jennifer A. Johnson, Melissa Ness, Dan Zucker, and Victor Silva Aguirre for 
good discussions.

S.F. was supported by the project grant "The New Milky" from the 
Knut and Alice Wallenberg foundation (PI: Feltzing).

\bibliography{SFeltzing_references}  

\end{document}